\documentclass[12pt]{iopart}
\usepackage{bm}
\usepackage{latexsym}
\usepackage{dcolumn}
\usepackage{iopams}
\usepackage{graphicx}
\usepackage{epsfig}
\usepackage{psfrag}
\usepackage{enumerate}
\def\be {\begin{equation}}
\def\ee {\end{equation}}
\def\bea {\begin{eqnarray}}
\def\eea {\end{eqnarray}}
\def\bse {\begin{subequations}}
\def\ese {\end{subequations}}
\def\bc {\begin{center}}
\def\ec {\end{center}}
\def\bfg {\begin{figure}}
\def\efg {\end{figure}}
\def\bi {\begin{itemize}}
\def\ei {\end{itemize}}
\def\bed {\begin{description}}
\def\eed {\end{description}}
\def\ben {\begin{enumerate}}
\def\een {\end{enumerate}}
\def\nn {\nonumber}

\def\la {\label}
\def\le {\left}
\def\ri {\right}
\def\pa {\partial}
\def\fr {\frac}
\def\sq {\sqrt}
\def\no {\noindent}

\def\ul {\underline}

\def\lra {\longrightarrow}

\def\a  {\alpha}
\def\b  {\beta}
\def\c  {\gamma}
\def\C  {\Gamma}
\def\d  {\delta}
\def\D  {\Delta}
\def\e  {\epsilon}

\def\l  {\lambda}

\def\s  {\sigma}
\def\t  {\tau}
\def\vph {\varphi}
\def\vth {\vartheta}

\def\sA {{\mbox{\tiny{\it A}}}}
\def\sB {{\mbox{\tiny{\it B}}}}
\def\sC {{\mbox{\tiny{\it C}}}}
\def\sD {{\mbox{\tiny{\it D}}}}

\def\sL {{\mbox{\tiny{\it L}}}}
\def\sM {{\mbox{\tiny{\it M}}}}
\def\sN {{\mbox{\tiny{\it N}}}}

\def\cA {\mathcal A}
\def\cB {\mathcal B}

\def\cM {\mathcal M}
\def\cU {\mathcal U}
\def\bcM {\overline{\mathcal M}}
\def\bcU {\overline{\mathcal U}}
\def\hcU {\widehat{\mathcal U}}
\def\fL {\mathfrak L}

\def\ba {\overline{a}}
\def\bb {\overline{b}}
\def\bc {\overline{c}}

\def\bg {\overline{g}}

\def\bp {\overline{p}}
\def\bq {\overline{q}}

\def\bu {\overline{u}}
\def\bR {\overline{R}}
\def\bU {\overline{U}}

\def\Rt {\widetilde{R}}
\def\Ct {\widetilde{\Gamma}}
\def\nab {\nabla\!}
\def\nt {\widetilde{\nabla}\!}

\long\def\symbolfootnote[#1]#2{\begingroup%
\def\thefootnote{\fnsymbol{footnote}}\footnote[#1]{#2}\endgroup}
\begin{document}

\title{Constraining Torsion in Maximally Symmetric (sub)spaces}

\author{Sourav Sur and Arshdeep Singh Bhatia}

\address{Department of Physics \& Astrophysics\\
University of Delhi, New Delhi - 110 007, India}

\eads{sourav.sur@gmail.com and asbhatia@physics.du.ac.in}

\begin{abstract}
We look into the general aspects of space-time symmetries in presence 
of torsion, and how the latter is affected by such symmetries. Focusing
in particular to space-times which either exhibit maximal symmetry on
their own, or could be decomposed to maximally symmetric subspaces, we
work out the constraints on torsion in two different theoretical schemes. 
We show that at least for a completely antisymmetric torsion tensor (for 
example the one motivated from string theory), an equivalence is set 
between these two schemes, as the non-vanishing independent torsion 
tensor components turn out to be the same. 
\end{abstract}

\pacs{04.20.-q, 04.50.-h, 04.50.Kd}

\vskip 0.25in
\noindent
{\it Keywords:} Torsion, Maximal symmetry, Riemann-Cartan space-time.

\maketitle

%\tableofcontents

%%%%%%%%%%%%%%%%%%%%%%%%%%%%%%%%%%%%%%%%%%%%%%%%%%%%%%%

\section{Introduction       \la{sec:intro}}

\no 
The geometry of Riemann-Cartan (RC) space-time has been of some importance both 
in the context of local field theories and in the effective scenarios originating 
from string theory. Such a space-time is characterized by an asymmetric (but 
metric-compatible) affine connection, the antisymmetrization of which in at 
least two of its indices gives rise to a third rank tensor field known as 
{\em torsion} \cite{hehl,akr,traut,sab,shap}. The incorporation of torsion 
is a natural modification of General Relativity (GR), especially from the point 
of view that at the simplest level a classical background can be provided for 
quantized matter fields (with definite spin), typically below the Planck scale
\cite{sab,shap,ham}. 

There are several predictions of the observable effects of torsion, whose 
origin could be traced from various sources. For instance, the massless 
antisymmetric Kalb-Ramond (KR) field in closed string theory has been argued
to be the source of a completely antisymmetric torsion in the background 
\cite{pmssg,saa}. Such a torsion has its significance in the effects produced 
on a number of astrophysical phenomena, which have been explored in detail 
both in the usual four dimensional framework and in the context of compact 
extra dimensional theories \cite{ssgss,skpm,dmpmssg,acpm,bmssgsen,ssgss1}. 
Investigations have also been carried out for the observable effects of 
gravitational parity violation, which is shown to be plausible in presence 
of torsion \cite{bmssgss}. Another subject of some relevance has been the 
study of torsion in local conformal frames, particularly in the context of 
non-minimal metric-scalar theories of gravitation \cite{shap,neto}. Certain 
non-minimal scalar-torsion couplings have been proposed \cite{shap,neto}, 
which may assign scalar field sources to some of the irreducible torsion 
modes and thereby produce interesting physical effects \cite{ss1,ssasb}. 
Non-minimal couplings of torsion with spinor fields are also of some importance, 
as specific bounds on certain torsion components for such couplings have been 
extracted from modern experimental searches for Lorentz violation \cite{russ}. 
In recent years, with major advances in the formal aspects of the Poincar\'e 
gauge theory of gravity \cite{PGT}, the effects of the associated torsion modes 
have been explored in the context of inflationary cosmology \cite{yonest} as 
well as for the problem of dark energy in the universe \cite{mink,shie}. Moreover, 
certain modified versions of the {\em teleparallel} (torsion without curvature) 
theories, known as $f(T)$ theories, are found to have interesting implications 
in cosmology and astrophysics \cite{capo1,flan,ior,boem1,barrow}.

Now, it is crucial to emphasize here that the estimation of any of the observable 
effects of torsion is subject to a clear understanding of how torsion is affected 
by, and in turn does affect, the symmetries of space-time. In other words, given 
a space-time metric structure, the foremost requirement for any observable 
prediction of torsion is the complete determination of the admissible torsion 
degrees of freedom, depending on the symmetries that are exhibited. Let us limit 
our attention to the scenarios of natural interest in any gravitational theory, 
viz. to space-times which either exhibit {\em maximal symmetry} on their own or 
could be decomposed to maximally symmetric subspaces. If the maximal symmetry is
to be preserved in presence of torsion, then one has to ascertain the existent 
torsion tensor components, taking note of the fact that these components would 
back-react on the metric and hence affect the geometric structure of space-time. 

One may recall that in GR, a tensor (of specific rank and symmetry properties) 
can in principle be constrained if it is invariant in form under the infinitesimal 
isometries of the metric of a given space. On the other hand, an $n$-dimensional 
space is said to be maximally symmetric if its metric admits the maximum number 
($n (n+1)/2$) of independent Killing vectors. Form-invariance of a tensor under 
maximal symmetry therefore implies the vanishing Lie derivative of the tensor 
with respect to each of these $n (n+1)/2$ Killing vectors. In presence of torsion 
however, one needs to be careful in dealing with the concepts of space-time 
symmetries and in particular the maximal symmetry. In fact, the geometric nature 
of torsion may reveal in the form-invariance of the torsion tensor with respect 
to all isometries if there is a mathematical principle which entails the space-time 
to have a definite structure, viz. either maximally symmetric in entirety or 
could be decomposed into maximally symmetric subspaces. Refer, for example, to 
the (large scale) homogeneity and isotropy of the universe --- the so-called 
{\em cosmological principle}. If such a principle is to be obeyed in presence 
of torsion then the entire four dimensional (RC) manifold should consist of 
space-like three dimensional maximally symmetric subspaces identified as 
hypersurfaces of constant cosmic time, and like any other cosmic tensor field 
torsion may be form-invariant under the isometries of the metric of such 
subspaces \cite{tsim}. This is logical at least when the torsion modes are 
specified in terms of some other degrees of freedom in the theory, for e.g. 
a scalar potential \cite{hoj,fneto,flan} or a second rank tensor potential
\cite{ham,pmssg,klein,gruv}. However, for the theories of rather conventional 
type, in which torsion is an independent variable that contributes either 
algebraically to the action (as in the original Einstein-Cartan formulation 
\cite{hehl,akr,sab}) or is propagating \cite{saa,sez,carroll1}, there is no 
concrete reason in support of the form-invariance of torsion. In such cases, 
the fundamental question: 
\bi 
\item {\em What is the precise meaning of maximal symmetry in presence of 
torsion?} 
\ei
has its relevance to the more appropriate question in all possible circumstances 
in which definite space-time structures involving torsion have to be maintained: 
\bi 
\item {\em How is the torsion tensor constrained either by virtue of its 
form-invariance under maximal symmetry, or in the course of defining maximal 
symmetry in its presence?}
\ei
From the technical point of view, the (manifestly covariant) Killing equation 
is given by the vanishing anticommutator of the covariant derivatives of the 
Killing vectors. In GR, this equation directly follows from the isometry 
condition, viz. vanishing Lie derivative of the metric tensor with respect to 
the Killing vectors \cite{wein}. In the RC space-time however, this equivalence 
no longer exists in general, as the covariant derivatives now involve 
torsion. Moreover, the equations relevant for the integrability of the Killing
equation\symbolfootnote[6]{In GR, the successive application of two 
equations satisfied by the curvature tensor (viz. its definition in terms of 
the anticommutator of two covariant derivatives of a vector, and its cyclicity 
property) actually leads to the integrability criterion, from which it is 
inferred that an $n$-dimensional maximally symmetric space has the maximum 
number $n(n+1)/2$ of Killing vectors.} are also manifestly covariant and hence 
get altered when expressed in terms of the covariant derivatives involving 
torsion in the RC space-time. So, one needs to sort out whether the preservation 
of the Killing equation or(and) the Killing integrability criterion is(are) 
absolutely necessary in space-times with torsion. If so, then what are the 
constraints on the torsion tensor? What are the constraints otherwise, under 
the demand that torsion is form-invariant in maximally symmetric spaces? 

A study of literature reveals that one can in principle resort to two different 
schemes (from contrasting viewpoints) in order to access the underlying aspects
of space-time symmetries (and in particular, of maximal symmetry) in presence 
of torsion \cite{tsim,zec,boem,bloom,dgssg,mult}:
\bed 
\item {\em Scheme I:} From a somewhat {\em weaker} viewpoint, maximal symmetry 
is to be understood solely from the metric properties of space-time. Therefore, 
in presence of torsion a maximally symmetric $n$-dimensional space is still the 
one which admits the maximum number $n(n+1)/2$ of Killing vectors, the latter 
satisfying the usual (general relativistic) Killing equation and the equations 
relevant for its integrability. However, as torsion is a characteristic of 
space-time, maximal symmetry has its significance only when it leaves torsion 
form-invariant --- a condition that imposes constraints on the torsion tensor, 
just as it would on any other third rank tensor (with the specific antisymmetry
property similar to torsion) \cite{tsim,zec,boem}. 
\item {\em Scheme II:} From a {\em stronger} viewpoint, maximal symmetry has 
to be understood in presence of torsion by explicitly taking into account 
torsion's effect on the Killing and other relevant equations, and demanding 
that the form of these equations should remain intact. This would however 
constrain the torsion tensor itself so that a maximally symmetric 
$n$-dimensional space (exhibiting torsion) would not only be the space which 
admits the maximum number of Killing vectors, but also that this maximum 
number would precisely be $n(n+1)/2$ (as in GR) \cite{bloom,dgssg,mult}.  
\eed
Either of these schemes may be useful for a self-consistent implementation 
of the concept of maximal symmetry in presence of torsion. However, the 
the scheme I, which actually suppresses the influence of torsion on such 
symmetry, is primarily applicable for torsion modes that are derived from 
some other degrees of freedom in the theory (as is common, for e.g., in 
many effective scenarios originating from string theory). As such, from 
the phenomenological point of view the scheme I is generally favoured 
in spite of the fact that one has to comply with the lack of appropriate 
covariant generalization of the equations of relevance (the Killing equation 
inclusive) in presence of torsion \cite{ss1,ssasb,tsim,zec,boem}. Scheme II, 
on the other hand, gives a clearer geometric picture and wider applicability,
but at the same time seems a bit too idealistic as it requires stringent 
conditions on the torsion tensor for apparently no reason other than to 
define maximal symmetry in the exact analogy of that in 
GR\symbolfootnote[7]{Moreover, a strict enforcement of both the minimal 
coupling prescription and the principle of general covariance is implied 
in the scheme II. Minimal coupling ensures that in presence of torsion the 
Riemannian covariant derivatives get replaced by the RC ones, so that all 
equations that involve covariant derivatives of tensors (of rank $\geq 1$) 
are in general altered. General covariance, on the other hand, suggests that 
in order to understand maximal symmetry one needs to take into account only 
the (altered) Killing and other relevant equations, as they are manifestly 
covariant in the RC space-time.} \cite{bloom,dgssg,mult}. Of course, the 
ambiguity in choosing which of the two schemes to follow could be resolved 
if it turns out that the outcome (in the form of the set of constraints on 
torsion) is the same. 

The objective of this paper is to make a systematic study of the independent, 
non-vanishing torsion tensor components in a manifold of given dimensionality, 
say $d$, which is either entirely maximally symmetric or consist of maximally 
symmetric subspaces of dimensionality  $n \, (< d)$. A major portion of the 
paper deals with the scheme I mentioned above. The analysis is carried out 
along the lines of Tsimparlis \cite{tsim}, in which the relevant components 
of the torsion tensor have been found under the demand of its form-invariance 
in a maximally symmetric subspace of dimensionality $n = 3$ or more, with 
emphasis on the homogeneous and isotropic cosmologies. We extend Tsimparlis' 
work to include all possible scenarios ($n \geq 2$) and also find the existent 
irreducible modes of torstion, viz. the trace, the completely antisymmetric 
(pseudo-trace) part, and the (pseudo)trace-free part, in a $d$-dimensional 
space-time. The remainder of our paper determines the constraints on the 
torsion tensor under the scheme II. Although our initial approach is similar 
to that in \cite{dgssg}, while clarifying the meaning of the maximal symmetry 
we concentrate only on the essential restrictions on torsion. As such, the 
constraints we find for the scheme II are in general different from those in 
\cite{dgssg}. We also make a comparison of the outcome of the schemes I and 
II, in order to establish a correlation, and possibly an equivalence between 
them. We actually observe that such an equivalence exists at least in the
case of a completely antisymmetric torsion tensor.   

The organization of the paper is as follows: in section \ref{sec:Tsym}, we 
discuss the basic concepts related to space-time symmetries (viz. isometries, 
Killing vectors, etc.) and how these concepts could possibly be understood 
in space-times admitting torsion. In section \ref{sec:App1}, taking the 
approach based on the scheme I we determine the existent components of the 
torsion tensor and its irreducible modes in a $d$-dimensional space-time 
manifold $\cM$ with a maximally symmetric sub-manifold $\bcM$ of dimension 
$n \,(< d)$. In section \ref{sec:App2}, we look for the constraints imposed 
on torsion in the formal development of theories based on the scheme II. A 
comparative study of the results obtained in the two schemes is done in 
section \ref{sec:Compare}, by resorting to certain scenarios of physical 
relevance. We conclude with a summary and some open questions in section 
\ref{sec:Concl}. The general aspects of a $d$-dimensional RC space-time, 
viz. the definition of torsion, its irreducible modes, the covariant 
derivatives, geodesics and auto-parallels, etc. are reviewed in the Appendix.

\section{Symmetric (sub)spaces and torsion \la{sec:Tsym}}

\no
Let us look into the basic concepts related to symmetric (sub)spaces and the 
role of torsion in influencing or preserving the symmetries. The starting 
point is the condition 
\be \la{genisom}
g_{\sA\sB} (x) ~=~ \fr{\pa x'^\sM}{\pa x^\sA} \fr{\pa x'^\sN}{\pa x^\sB} 
\, g_{\sM\sN} (x') \,\,,
\ee
from the requirement that certain coordinate transformations $x \lra x'$, 
known as {\em isometries}, would leave the metric tensor form-invariant, 
$g'_{\sA\sB} (x') = g_{\sA\sB} (x')$ \cite{wein}. For infinitesimal 
isometries, viz. $x'^\sM = x^\sM + \xi^\sM (x)$, the above condition 
(\ref{genisom}) reads (to the first order in $\xi^\sM$ and its derivatives): 
\be \la{infisom}
\fL_\xi \, g_{\sA\sB} (x) ~=~ 0 ~=~ \xi^\sM \, \pa_\sM \, g_{\sA\sB} ~+~ 
g_{\sM\sB} \, \pa_\sA \xi^\sM ~+~ g_{\sA\sN} \, \pa_\sB \xi^\sN \,\,,
\ee
where $\fL_\xi$ denotes the Lie derivative with respect to the vector 
$\xi^\sM$. In a similar manner, form invariance of any tensor, for e.g. 
torsion $T_{\sA \sB \sC}$, under the infinitesimal isometries of the metric 
would imply the vanishing Lie derivative of the tensor with respect 
to $\xi^\sM$, i.e. $\fL_\xi \, T_{\sA \sB \sC} \,=\, 0$. Eq. (\ref{infisom}) 
is of crucial importance in GR for understanding the aspects of symmetries, 
and in particular the maximal symmetry of a given space or subspaces.

\subsection{Maximal symmetry in absence of torsion \la{sec:Rsym}}

\no
In the Riemannian space-time (without torsion), Eq. (\ref{infisom}) can 
be expressed in a covariant form --- the so-called {\em Killing equation} 
in GR:
\be \la{R-Killeq}
\nab_\sA \, \xi_\sB ~+~ \nab_\sB \, \xi_\sA ~=~ 0 \,,
\ee
and any vector field which satisfies this equation is said to form a {\em 
Killing vector} of the metric $g_{\sA\sB} (x)$. Hence the infinitesimal 
isometries of the metric are essentially determined by the space of vector 
fields spanned by the Killing vectors \cite{wein}. Moreover, when a metric 
space admits the maximum possible number of linearly independent Killing 
vectors, the space is said to be {\em maximally symmetric}. Now, what is 
this maximum possible number? To determine this, one uses the following 
two relations in GR:  
\ben[(i)]
\item the commutator of two covariant derivatives of a vector in terms 
of the product of the vector and the Riemann curvature tensor:
\be \la{R-commvec}
\le(\nab_\sA \nab_\sB ~-~ \nab_\sB \nab_\sA\ri) \xi_\sC ~=~ 
- \, R^\sM_{~\sC\sA\sB} \, \xi_\sM \,\,,  
\ee
\item the cyclic sum rule for the Riemann curvature tensor:
\be \la{R-cycl}
R^\sM_{~\sA\sB\sC} ~+~ R^\sM_{~\sB\sC\sA} 
~+~ R^\sM_{~\sC\sA\sB} ~=~ 0 \,\,.
\ee
\een
Adding with Eq. (\ref{R-commvec}) its two cyclic permutations 
in the indices $A, B, C$, and using Eqs. (\ref{R-cycl}) and 
(\ref{R-Killeq}), we get 
\be \la{R-intcond1}
\nab_\sA \nab_\sB \, \xi_\sC ~-~ \nab_\sB \nab_\sA \, \xi_\sC ~-~ 
\nab_\sC \nab_\sB \, \xi_\sA ~=~ 0 \,\,,
\ee
whence the Eq. (\ref{R-commvec}) becomes
\be \la{R-intcond}
\nab_\sC \nab_\sB \, \xi_\sA ~=~ -\, R^\sM_{~\sC\sA\sB} \, \xi_\sM \,\,.
\ee
This is the {\em integrability condition} for the Killing vectors, which 
implies that any particular Killing vector $\xi_\sM (x)$ of the metric 
$g_{\sA\sB} (x)$ is specified uniquely by the values of the Killing vector 
and its covariant derivative at any particular point $X$, i.e. by 
$\xi_\sM (X)$ and $\nab_\sN \xi_\sM \vert_{x=X}$. As a result, in an 
$n$-dimensional metric space there can be at most $n (n+1)/2$ Killing 
vectors $\xi_\sM^{(q)} (x)$ which do not satisfy any linear relation of 
the form $\sum_q c_q \, \xi_\sM^{(q)} (x) = 0$, with constant coefficients 
$c_q$ \cite{wein}.

For spaces with maximally symmetric subspaces, the analysis is similar to 
the above. However, the maximum number of independent and non-vanishing Killing
vectors that are admitted for a family of such subspaces, say of dimensionality
$n$, are only $n(n+1)/2$, and not $d(d+1)$, if  the entire space-time is of
dimensionality $d$. Hence, the constraints on a tensor due to its form-invariance 
under the infinitesimal isometries of the metric, in the case of spaces with 
maximally symmetric subspaces, would in general be different from those in the 
scenario where the entire space-time is maximally symmetric. We shall look into 
this rather explicitly in what follows. However, for the time being, let us first 
concentrate on how to go about understanding maximal symmetry when the space-time 
admits torsion, in the next subsection.

\subsection{Maximal symmetry in presence of torsion \la{sec:RCsym}}

\no
Torsion being a geometric entity, it is fair to argue that the form-invariance 
of the torsion tensor (in addition to that of the metric) is a requirement for 
the preservation of maximal symmetry. But how do we conceive maximal symmetry 
afterall, in presence of torsion? The question amounts to justify which should 
be taken to be fundamental --- the equation (\ref{infisom}) giving the condition 
of form-invariance of the metric tensor under infinitesimal isometries, or the 
explicitly covariant Killing equation (\ref{R-Killeq}) which is in general 
modified in presence of torsion (and so are the other relevant equations 
(\ref{R-commvec}) and (\ref{R-cycl})). As mentioned earlier, there are two 
different schemes for the implementation of maximal symmetry in space-times 
with torsion:
\bi
\item The one which sets aside the general covariance (and also minimal coupling), 
and considers Eq. (\ref{infisom}) to be the most fundamental, as this follows 
straightaway from the first principles. Maximal symmetry is then to be realized 
precisely in the same way as in GR, i.e. the analysis in the previous subsection 
would go through even in presence of torsion. The only objective that remains is 
to determine the constraints on the torsion tensor on account of its form-invariance 
under the maximal symmetry.
\item The other which takes the general covariance (and the minimal coupling) in 
a serious note, and hence considers the modified version of Eq. (\ref{R-Killeq}), 
and of the follow-up Eqs. (\ref{R-commvec}) and (\ref{R-cycl}) in torsioned spaces, 
to be fundamental. In fact, the demand is to be that all these equations should 
retain their forms, even when the covariant derivatives involve torsion (i.e.
$\nab_\sA$ is replaced by $\nt_\sA$, see the Appendix for notations and definitions). 
The analysis in the previous subsection would then again go through, however at 
the expense of constraining the torsion tensor severely for such restoration of 
forms of the above equations. 
\ei
In the next two sections we work out the independent non-vanishing components of 
the torsion tensor (and also of its irreducible modes), by taking into account 
one-by-one the constraining equations on torsion in the above two schemes. 
Thereafter, resorting to some specific scenarios of physical importance, we make 
a comparison of the allowed torsion degrees of freedom in these two schemes. In
particular, we look for the cases in which the independent torsion components 
allowed by the two schemes turn out to be the same. That would at least partially 
resolve the somewhat conflicting issue of maximal symmetry in a given theory 
involving torsion.

\section{Scheme I : Constraints on a maximally form-invariant torsion  
\la{sec:App1}}

\no
In this section we consider the usual (general relativistic) definition of 
maximal symmetry in presence of torsion, and work out the constraints on 
the torsion tensor due to its form-invariance under such symmetry. This is 
the scheme I mentioned above, in which the fundamental aspects of symmetries 
of metric spaces are supposedly governed by Eq. (\ref{infisom}) that follows 
from the first principles. We proceed along the lines of Tsimparlis \cite{tsim} 
to determine the non-vanishing independent components of the torsion tensor 
in maximally symmetric (sub)spaces. Our notations and conventions are as 
follows:  
\bi
\item $\cM$ is a $d$-dimensional (``bulk") manifold, with a non-degenerate 
symmetric metric $g_{\sA\sB} (x)$, where $x :\equiv \{x^\sA\}$ are the bulk 
coordinates and $A,B,\dots$ are the bulk indices (each of which runs over 
all the $d$ labels $0,1,2,\dots,d-1$).
\item $\bcM$ is a maximally symmetric $n$-dimensional submanifold of $\cM$, 
with metric $\bg_{\ba\bb} (\bu)$, where $\bu :\equiv \{\bu^{\ba} \}$ are 
the coordinate labels in $\bcM$, and each of the corresponding indices 
$\ba,\bb,\dots$ runs over $n$ of the $d$ labels.
\item The quotient $\cM/\bcM$ is in general not maximally symmetric, and is 
just a topological space, with topology induced from $\cM$. Although not 
necessarily a manifold, we assume $\cM/\bcM$ to be such (but in general not 
a submanifold of $\cM$), with metric $g_{ab} (v)$, where $v :\equiv \{v^a \}$ 
are the coordinate labels, and each of the indices $a,b,\dots$ runs over the 
remaining $(d\!-\!n)$ of the $d$ labels.
\ei
In fact, the assumption that $\cM/\bcM$ is a manifold amounts to mentioning 
that we are restricting our analysis to an open set $\cU \subset \cM$, which
is diffeomorphic to a direct product
\be \la{subset}
\cU \,=\, \bcU \times \hcU \,\,, \qquad \bcU \subset \bcM \,, \qquad 
\hcU \subset \cM/\bcM \,,     
\ee
where $\bcU$ and $\hcU$ are also open sets. We then choose local coordinates
adapted to this diffeomorphism, and split the coordinates on $\cM$ as \,
$x^\sA := \le(\bu^{\ba}, v^a\ri)$. This is reasonable as the graviatational
field equations are of local nature. One solves them in open sets and then,
through a mechanism of gluing (if necessary), they can be extended to other,
intersecting open sets to construct a global manifold.

We shall look into all possible scenarios $2 \leq n \leq d$, the special case 
$n=d$ of course deals with a maximally symmetric bulk manifold $\cM$. For 
\,$n \neq d$\, however, we shall resort to a rather complicated picture, viz. 
that of the $d$-dimensional manifold $\cM$ being decomposed in a family of 
maximally symmetric $n$-dimensional sub-manifolds $\bcM$. Then the above splitting
$x^\sA := \le(\bu^{\ba}, v^a\ri)$ of the bulk coordinates has a succinct implication, 
and there are stringent conditions on the bulk metric $g_{\sA\sB} (x)$ \cite{wein}:
\bi
\item The sub-manifold $\cM$, the coordinate functions on which are $\bu^{\ba}$ with 
$\ba$ taking $n$ of the $d$ values $0,1,2,\dots,d-1$, are distinguished from one
another by the coordinate labels $v^a$, where the index $a$ can take the remaining
$(d-n)$ of the $d$ values.
\item The subspaces with constant $v^a$ are maximally symmetric subject to the condition
that the bulk metric $g_{\sA\sB} (x)$ is invariant under the infinitesimal transformations
\be \la{inftrans}
\bu^{\ba} ~\lra~ \bu'^{\ba} ~=~ \bu^{\ba} ~+~ \xi^{\ba} (\bu,v) \quad , 
\qquad v^a ~\lra~ v'^a ~=~ v^a \,\,,
\ee
where $\xi^{\ba}$ are the Killing vectors in $\cM$. There are \,$n(n+1)/2$\, such Killing 
vectors $\xi^{\ba}$ which are linearly independent.  As the transformations (\ref{inftrans})
leave the coordinates $v^a$ invariant, the Killing vectors $\xi^a (\bu,v)$ are all zero. 
\item Finally, there is a powerful theorem which states that: it is always possible to 
choose the coordinates $\bu^{\ba}$ such that the bulk metric $g_{\sA\sB}$ 
is decomposed as 
\be \la{bulkmet}
g_{\sA\sB} \, dx^\sA \, dx^\sB ~=~ g_{ab} (v) \, dv^a \, dv^b ~+~ 
f (v) \, \bg_{\ba\bb} (\bu) \, d\bu^{\ba} \, d\bu^{\bb} \,\,,
\ee
where $f (v)$ is some specific function of the $v$-coordinates only. 
Eq. (\ref{bulkmet}) implies that there are no mixed elements of the 
form $g_{\ba a}$, and the Killing vectors in $\bcM$ do not depend on 
the $v$-coordinates of $\cM/\bcM$, i.e. $\xi^{\ba} = \xi^{\ba} (\bu)$. 
\ei
Now, the form-invariance of torsion under the isometries of the metric 
in $\bcM$ is given by the relation: 
\be \la{torinv}
\fL_{\xi} \, T_{\sA \sB \sC} \,=\, 0 \,=\, 
\xi^{\bq} \, \pa_{\bq} \, T_{\sA \sB \sC} \,+\, 
T_{\bq \sB \sC} \, \pa_\sA \xi^{\bq} \,+\, 
T_{\sA \bq \sC} \, \pa_\sB \xi^{\bq} \,+\, 
T_{\sA \sB \bq} \, \pa_\sC \xi^{\bq} \,.
\ee
On the other hand, the maximal symmetry in $\bcM$ implies that a 
Killing vector $\xi^{\bq}$ be so chosen that 
\ben[(i)]
\item it vanishes at any given point $\bU$ in $\bcM$, i.e. \,$\xi^{\bq} 
(\bU) = 0$,\, whereas 
\item its covariant derivative (defined in terms of the Christoffel 
connection) at that point, i.e. \,$\nab_{\bp} \,\xi_{\ba} \vert_{\bu = 
\bU}$,\, forms an arbitrary matrix, which is of course antisymmetric 
because of the Killing equation ~\,$\nab_{\bb} \, \xi_{\ba} \,+\, 
\nab_{\ba} \, \xi_{\bb} \,=\, 0$.
\een
Therefore at $\bu = \bU$, Eq. (\ref{torinv}) gives
\be \la{torinv1}
\le(\d_{\sA}^{\bp} \, T^{\bq}_{~\sB\sC} ~+~ 
\d_{\sB}^{\bp} \, T^{~\bq}_{\sA~\sC} ~+~
\d_{\sC}^{\bp} \, T^{~~\,\bq}_{\sA\sB}\ri) \nab_{\bp} \, \xi_{\bq} 
~=~ 0 \,\,,
\ee
and $\nab_{\bp} \, \xi_{\bq}$ being arbitrary and antisymmetric, its 
coefficient is symmetric in $\bp$ and $\bq$:
\be \la{torinv2}
\d_{\sA}^{\bp} \, T^{\bq}_{~\sB\sC} ~+~ 
\d_{\sB}^{\bp} \, T^{~\bq}_{\sA~\sC} ~+~
\d_{\sC}^{\bp} \, T^{~~\,\bq}_{\sA\sB} ~=~ 
\d_{\sA}^{\bq} \, T^{\bp}_{~\sB\sC} ~+~ 
\d_{\sB}^{\bq} \, T^{~\bp}_{\sA~\sC} ~+~ 
\d_{\sC}^{\bq} \, T^{~~\,\bp}_{\sA\sB} \,\,.
\ee
This equation should hold everywhere, since the point $\bU$ is 
arbitrary as well.

\subsection{Allowed independent components of the torsion 
tensor \la{sec:indtor}}

\no
In view of the antisymmetry property of torsion, viz. $T^\sA_{~\sB\sC} 
= - T^\sA_{~\sC\sB}$, we have the following six conditions \cite{tsim}:
\bea \la{torinvall}
&\,\, \mbox{(i)} \,\, \fL_{\xi^{\bq}} \, T_{\ba \bb \bc} ~=~ 0 \,\,, \quad 
\mbox{(ii)} \,\, \fL_{\xi^{\bq}} \, T_{\ba \bb a} ~=~ 0 \,\,, \quad
\mbox{(iii)} \,\, \fL_{\xi^{\bq}} \, T_{\ba a b} ~=~ 0 \,\,, \nn\\
&\mbox{(iv)} \,\, \fL_{\xi^{\bq}} \, T_{a b \ba} ~=~ 0 \,\,, \quad 
\mbox{(v)} \,\, \fL_{\xi^{\bq}} \, T_{a \ba \bb} ~=~ 0 \,\,, \quad
\mbox{(vi)} \,\, \fL_{\xi^{\bq}} \, T_{a b c} ~=~ 0 \,\,.
\eea
Let us now examine the outcome of these conditions in detail.
\bi 
\item For the condition (i), Eq. (\ref{torinv2}) leads to
\be \la{cond1}
\d^{\bp}_{\ba} \, T^{\bq}_{~\,\bb \bc} ~+~ 
\d^{\bp}_{\bb} \, T^{~\bq}_{\ba~\bc} ~+~ 
\d^{\bp}_{\bc} \, T^{~~\,\bq}_{\ba\bb} ~=~ 
\d^{\bq}_{\ba} \, T^{\bp}_{~\,\bb \bc} ~+~ 
\d^{\bq}_{\bb} \, T^{~\bp}_{\ba~\bc} ~+~ 
\d^{\bq}_{\bc} \, T^{~~\,\bp}_{\ba\bb} \,\,.
\ee
Contracting $\bp$ with $\ba$ gives
\be \la{cond1a}
\le(n - 1\ri) T^{\bq}_{~\, \bb\bc} ~+~ 
T^{~\bq}_{\bb~\bc} ~+~ T^{~~\, \bq}_{\bc\bb} ~=~ 
\d^{\bq}_{\bb} \, T^{~\ba}_{\ba~\bc} ~+~ 
\d^{\bq}_{\bc} \, T^{~~\, \ba}_{\ba\bb} \,\,.
\ee
Contracting further $\bq$ with $\bb$ results in 
$T^{\ba}_{~\, \ba\bc} = 0$, which when substituted back in 
Eq. (\ref{cond1a}) yields finally
\be \la{fincond1}
T_{\ba\bb\bc} =
\cases{
      T_{[\ba\bb\bc]} (v)  &, \quad $n = 3 \leq d$ \\
      0   & , \quad $n \neq 3$   
} \,\,.
\ee

\item For the condition (ii), Eq. (\ref{torinv2}) gives
\be \la{cond2}
\d^{\bp}_{\ba} \, T^{\bq}_{~\,\bb a} ~+~ 
\d^{\bp}_{\bb} \, T^{~\bq}_{\ba~a} ~=~ 
\d^{\bq}_{\ba} \, T^{\bp}_{~\,\bb a} ~+~ 
\d^{\bq}_{\bb} \, T^{~\bp}_{\ba~a}  \,\,.
\ee
Contracting $\bp$ with $\ba$ we get
\be \la{cond2a}
\le(n - 1\ri) T_{\bq\bb a} ~=~ - \, T_{\bb\bq a} ~+~ 
\d_{\bq\bb} \, T^{\ba}_{~\, \ba a} \,\,.
\ee
Now interchanging $\bq$ and $\bb$, then multiplying both sides by 
$(n-1)$, and subtracting the resulting equation from Eq. (\ref{cond2a}), 
one finally concludes
\bea \la{fincond2}
T_{\ba\bb a} = - T_{\ba a \bb} =
\cases{
      T_{[\ba\bb] a} (v) ~-~ \fr 1 2 \, \d_{\ba\bb} \, 
      T^{\bc}_{~\, a \bc} (v)  &, \quad $n = 2$ \\
      -\, \fr 1 n \, \d_{\ba\bb} \, T^{\bc}_{~\, a \bc} (v) 
       & , \quad $3 \leq n < d$   
} \,.
\eea

\item For the conditions (iii) and (iv), one gets in a similar manner 
the following
\be \la{fincond3-4}
T_{\ba a b} = 0 \quad , \qquad \mbox{and} \qquad T_{a b \ba} = 0 \quad , 
\quad (\forall ~ n \geq 2) \,,
\ee
whereas the condition (v) leads to
\bea \la{fincond5}
T_{a \ba\bb} = 
\cases{
      T_{a [\ba\bb]} (v) &, \quad $n = 2 < d$  \\
      0  &, \quad $n \neq 2$
} \,\,.
\eea

\item The condition (vi) is actually redundant, i.e. it does not provide
any additional information about the component $T_{a b c}$. However, 
since the indices $a, b, c$ are of the $(d-n)$-dimensional space $\cM/\bcM$ 
and $T_{a b c}$ is antisymmetric in $b$ and $c$, we have
\be \la{fincond6} 
T_{a b c} = T_{a [b c]} (v) \quad , \quad (\forall ~ n \leq d-2) \,\,. 
\ee
\ei
There is also a crucial point to note for the components $T_{\ba\bb a}$
and $T_{a \ba\bb}$ in the case $n=2$. Eq. (\ref{fincond2}) implies that 
$T_{\ba\bb a}$ is antisymmetric in $\ba$ and $\bb$ when these two indices 
are unequal (for $n=2$). But $T_{\ba\bb a}$ is antisymmetric in the last 
two indices as well. Therefore, the conclusion is that $T_{\ba\bb a}$ is 
completely antisymmetric, i.e. $T_{\ba\bb a} = T_{[\ba\bb a]}$ for $n=2$ 
if it is pre-assigned that $\ba \neq \bb$. Moreover, a completely 
antisymmetric $T_{\ba\bb a}$ is also equal to $T_{a \ba\bb}$. Thus to 
summarize, in a $d$-dimensional manifold $\cM$ the set of non-vanishing 
independent components of the torsion tensor $T_{\sA\sB\sC}$, that 
preserves the maximal symmetry of a sub-manifold $\bcM$ of dimensionality 
$n$, is given by
%
%\bse \la{indtorcomps}
\bea 
\la{indtorcomps1}
T_{\ba\bb\bc} &=& \e_{\ba\bb\bc} \, \b (v) \quad ; 
\hskip 1.3in (\mbox{only for} ~ n = 3 \leq d) \,\,,\\
\la{indtorcomps2}
T_{\ba\bb a} &=&  
\cases{
      T_{[\ba\bb a]} (v) ~-~ \fr 1 2 \, \d_{\ba\bb} \, \a_a (v) 
       &; \quad (\mbox{for} $n = 2 < d$) \\
      -\, \fr 1 n \, \d_{\ba\bb} \, \a_a (v) 
       &; \quad (\mbox{for} $3 \leq n < d$)   
} , \\
\la{indtorcomps3}
T_{a b c} &=& T_{a [b c]} (v) \quad ~; 
\hskip 1.3in (\mbox{for} ~ 2 \leq n \leq d-2) \,\,. 
\eea
%\ese
%
In the above, $\b (v)$ is a pseudo-scalar and $\a_a (v) = 
T^{\ba}_{~\, a \ba} (v)$ is a vector in the $(d\!-\!n)$-dimensional 
space $\cM/\bcM$. 

One may note that in the particular case where the entire bulk manifold 
$\cM$ is maximally symmetric (i.e. $n=d$), the torsion tensor can have 
only its completely antisymmetric part ~$T_{[\sA\sB\sC]} \equiv 
T_{[\ba\bb\bc]}$~ non-vanishing (and constant), and the dimensionality 
of the bulk can only be $d=3$.

\subsection{Allowed independent components of the irreducible torsion 
modes  \la{sec:tormodes}}

\no
Let us refer to Eq. (\ref{torirrd}) in the Appendix, for the 
irreducible decomposition of the torsion tensor $T_{\sA\sB\sC}$ in a 
$d$-dimensional space-time. The irreducible modes are the trace of 
torsion $T_\sA := T^\sB_{~\,\sA\sB}$, the totally antisymmetric 
part (or the pseudo-trace) $A_{\sA\sB\sC} := T_{[\sA\sB\sC]}$, 
and the (pseudo-)tracefree part $Q_{\sA\sB\sC}$ that satisfies 
the conditions (\ref{Q-tor}).

\subsubsection{Constraints on the trace of torsion \\ \la{sec:trace}}

\no
The components of the torsion trace vector $T_\sA$ are given by
\be \la{tracecomps}
T_{\ba} := T^\sA_{~\, \ba\sA} = T^{\bb}_{~\, \ba\bb} + T^a_{~\, \ba a} \, , 
\quad \mbox{and} \qquad
T_a := T^\sA_{~\, a \sA} = T^{\ba}_{~\, a \ba} + T^b_{~\, a b} \,\, .
\ee
Now, as shown above, $T_{\ba\bb\bc}$ is either completely antisymmetric 
(the case $n=3$) or zero (for $n \neq 3$). Therefore, its trace 
$T^{\bb}_{~\, \ba\bb} = 0 ~ (\forall n)$. Also, since $T_{a b \ba} = 0 
~ (\forall n)$, we have $T^a_{~\, \ba a} = 0 ~ (\forall n)$. Hence,
\be \la{tracecomp1}
T_{\ba} = 0 \quad, \qquad (\forall ~ n \geq 2) \,\,.
\ee
Similarly, it can be shown that
\bea \la{tracecomp2}
T_a =
\cases{
      \a_a (v) \quad &, \qquad  $n = d-1$ \\
      \a_a (v) ~+~ \c_a (v) \quad & , \qquad  $n < d-1$  
} \,\,,
\eea
where $\a_a \,=\, T^{\ba}_{~\, a \ba}$ (as before), and we have defined 
another vector $\c_a \,:=\, T^b_{~\, a b}$ in $\cM/\bcM$.

\subsubsection{Constraints on the completely antisymmetric part of torsion \\
\la{sec:pstrace}}

\no
The independent components of the tensor $A_{\sA\sB\sC}$ are: 
$\, A_{\ba\bb\bc} \,, \, A_{\ba\bb a} \,, \, A_{\ba a b} \,$ and 
$A_{a b c} \,$. An analysis similar to the above reveals that
the allowed ones are
\bea
A_{\ba\bb\bc} &=& \e_{\ba\bb\bc} \, \b (v) \quad ;  \qquad 
(\mbox{only for}~ n = 3 \leq d)  \,\,, \\  
\la{pstracecomp1}
A_{\ba\bb a} &=& T_{[\ba\bb a]}(v)  
\quad ; \qquad (\mbox{only for}~ n = 2 < d) \,\,, \\  
\la{pstracecomp2}
A_{a b c} &=& T_{[a b c]} (v) 
\quad ; \qquad ~ (\mbox{for} ~ 2 \leq n \leq d-3) \,\,,
\la{pstracecomp3}
\eea
where once again $\b$ is a pseudo-scalar that depends only on the coordinates
$v^a$ of $\cM/\bcM$.

\subsubsection{Constraints on the (pseudo-)tracefree part of torsion \\ 
\la{sec:tracefree}}

\no
The tensor $Q_{\sA\sB\sC}$ given by (see the Appendix)
\be \la{Q-tordef}
Q_{\sA\sB\sC} ~=~ T_{\sA\sB\sC} ~-~ \fr 1 {d-1} \le(g_{\sA\sC} \, T_\sB 
~-~ g_{\sA\sB} \, T_\sC\ri) \, -~ A_{\sA\sB\sC} \,\,,
\ee
satisfies the conditions (\ref{Q-tor}). In general, the independent components 
are: $Q_{\ba\bb\bc} \,,\, Q_{\ba\bb a} \,, \, Q_{\ba a b} \,, \, Q_{a b \ba} \,, 
\, Q_{a \ba\bb} \,$ and $Q_{a b c} \,$. However, the allowed ones in maximally 
symmetric (sub)spaces are the following:
\bea
Q_{\ba\bb a} &=& \d_{\ba\bb} \le[\le(\fr 1 {d-1} - \fr 1 n\ri) \a_a (v)
+ \le(\fr 1 {d-1}\ri) \c_a (v)\ri]  ; 
\quad (n < d-1) \,\,, \\  
\la{tracefreecomp1} 
Q_{a b c} &=& W_{a [b c]} (v) \hskip 2.45in ; \quad (\forall ~ n) \,,
\la{tracefreecomp2}
\eea
where 
\bea \la{W-def}
W_{a b c} &=& \fr 4 3 \, T_{(a b) c} \,+\, \fr 2 {d-1} \, g_{a b} \,T_c \\
&=& \cases{
\fr 2 {d-1} \, g_{a b} \, \a_c &, \, ($n = d-1$) \\
\fr 4 3 \, T_{(a b) c} \,+\, \fr 2 {d-1} g_{a b} \le[\a_c + \c_c\ri] &, \, ($n < d-1$)
} \,. \nn
\eea
%  

%\newpage

\section{Scheme II : Torsion in a generally covariant maximal symmetric 
set-up \la{sec:App2}}

\no
Let us now look into the concept of maximal symmetry in presence of torsion 
when the principle of general covariance is strictly obeyed. Under the 
minimal coupling prescription, the covariant derivatives of a $d$-dimensional 
Riemannian space (R$_d$) are generalized to those of a space admitting 
torsion (U$_d$), i.e. $\nab_\sA \lra \nt_\sA$. We demand that the Killing 
equation (\ref{R-Killeq}) should be preserved in form, when expressed in 
terms of these new covariant derivatives $\nt_\sA$, i.e.
\be \la{U-Killeq}
\nt_\sA \, \xi_\sB ~+~ \nt_\sB \, \xi_\sA ~=~ 0 \,\,. 
\ee
Such a demand is actually based on the argument that 
the Killing vectors are used to determine the constants associated with 
the motion along the affine curves with properties defined by the metric 
(or)and the connection. In the space-times admitting torsion such curves are  
the auto-parallels (sometimes called the affine geodesics) which transport
their tangent vectors parallely to themselves. These curves are in general 
different from the usual (metric) geodesics which extremize the separation 
between events and depend only on the metric properties of space-time (see 
the Appendix for details). If a vector $v^\sA = dx^\sA/d\s$ is tangent to 
an auto-parallel curve affinely parameterized by $\s$, then the constants
of motion are determined from \cite{poiss}
\be \la{const-motion}
\fr d {d\s} \le(v^\sA \, \xi_\sA\ri) =~ 0 \,.
\ee
However, in the U$_d$ space-time one has
\be \la{motion}
\fr d {d\s} \le(v^\sA \, \xi_\sA\ri) =~ 
v^\sB \, \nt_\sB \le(v^\sA \, \xi_\sA\ri) =~ 
\xi_\sA \, v^\sB \, \nt_\sB \, v^\sA ~+~ 
v^\sA \, v^\sB \, \nt_\sB \, \xi_\sA \,.
\ee
The first term on the right hand side of course vanishes by virtue of the
auto-parallel equation $v^\sB \nt_\sB v^\sA = 0$ [{\it cf.} Eq. (\ref{autop}) 
in the Appendix], but the second term would vanish only when we assert that 
the Killing vectors satisfy the above equation (\ref{U-Killeq}). Moreover,
since the Killing vectors also satisfy the relation $\nab_\sA \xi_\sB + 
\nab_\sB \xi_\sA = 0$  [{\it cf.} Eq. (\ref{R-Killeq})], they determine the 
constants of motion along the metric geodesics as well. That is to say, if 
$u^\sA = dx^\sA/d\l$ is tangent to a metric geodesic parameterized by $\l$ 
then we have  
\be \la{motion1}
\fr d {d\l} \le(u^\sA \, \xi_\sA\ri) =~ 
u^\sB \, \nab_\sB \le(u^\sA \, \xi_\sA\ri) =~ 
\xi_\sA \, u^\sB \, \nab_\sB \, u^\sA ~+~ 
u^\sA \, u^\sB \, \nab_\sB \, \xi_\sA ~=~ 0 \,,
\ee
as in Riemannian space-time. However, the parameter $\l$ may not be an affine
parameter in space-times involving torsion (except in the case of a completely 
antisymmetric torsion tensor for which the metric geodesics are identical
with the autoparallels, and one may verify that Eqs. (\ref{R-Killeq}) and
(\ref{U-Killeq}) are also the same)\symbolfootnote[2]{It is to be noted 
that for the scheme I, the Killing vectors can determine the constants of 
motion along the (metric) geodesics but not in general along the auto-parallels. 
So this scheme is primarily applicable when the torsion modes could be traded
away with some other fields in the theory (via constraint equations, as torsion 
is auxiliary). In such cases the Riemann-Cartan action effectively reduces to
the Riemannian one coupled with other fields.}.

Now, for the above Killing equation (\ref{U-Killeq}) to hold alongwith the 
Eq. (\ref{R-Killeq}), the torsion tensor should satisfy
\be \la{tor-Killcond}
\le(T_{\sA\sB\sC} ~+~ T_{\sB\sA\sC}\ri) \xi^\sC ~=~ 0 \,\,,
\ee
and if we proceed exactly as in GR (see section \ref{sec:Rsym}), we first 
encounter the equation
\be \la{U-commvec}
\le(\nt_\sA \nt_\sB ~-~ \nt_\sB \nt_\sA\ri) \xi_\sC ~=~ 
- \, \Rt^\sM_{~\sC\sA\sB} \, \xi_\sM ~+~ 
T^\sM_{~\sA\sB} \, \nt_\sM \, \xi_\sC \,\,,  
\ee
which is of course the generalization of Eq. (\ref{R-commvec}). Here, 
$\Rt^\sM_{~\sC\sA\sB}$ is the U$_d$ analogue of the Riemannian curvature 
tensor $R^\sM_{~\sC\sA\sB}$ :
\be \la{U-curvtens}
\Rt^\sM_{~\sC\sA\sB} ~=~ R^\sM_{~\sC\sA\sB} ~+~ 
\bR^\sM_{~\sC\sA\sB} \,\, ,  
\ee
where
\be \la{U-Rbar}
\bR^\sM_{~\sC\sA\sB} ~=~ 
\nab_\sA \, K^\sM_{~\sC\sB} ~-~ \nab_\sB \, K^\sM_{~\sC\sA} ~+~ 
K^\sM_{~\sN\sA} K^\sN_{~\sC\sB} ~-~ K^\sM_{~\sN\sB} K^\sN_{~\sC\sA} 
\,\,. 
\ee
Now, adding with Eq. (\ref{U-commvec}), its two cyclic permutations in the 
indices $A, B$ and $C$, then using the Killing equation (\ref{U-Killeq}) 
and the cyclicity condition (\ref{R-cycl}) for $R^\sA_{~\sB\sC\sD}$, we get 
an equation similar to Eq. (\ref{R-intcond1}) :
\be \la{U-intcond1}
\nt_\sA \nt_\sB \, \xi_\sC ~-~ \nt_\sB \nt_\sA \, \xi_\sC ~-~ 
\nt_\sC \nt_\sB \, \xi_\sA ~=~ 0 \,\,,
\ee
under the condition
\bea \la{tor-intcond}
\le(\bR^\sM_{~\sA\sB\sC} + \bR^\sM_{~\sB\sC\sA} + 
\bR^\sM_{~\sC\sA\sB}\ri) \xi_\sM ~= \nn\\
\hskip 1in - \le(T^\sM_{~\sA\sB} \, \nt_\sC \, \xi_\sM ~+~ 
T^\sM_{~\sB\sC} \, \nt_\sA \, \xi_\sM ~+~
T^\sM_{~\sC\sA} \, \nt_\sB \, \xi_\sM\ri) \,\,.
\eea
Substituting Eq. (\ref{U-intcond1}) back in Eq. (\ref{U-commvec}), and 
using the Killing equation (\ref{U-Killeq}) once more, we obtain
\be \la{U-intcond}
\nt_\sC \nt_\sB \, \xi_\sA ~=~ - \, \Rt^\sM_{~\sC\sA\sB} \, \xi_\sM ~-~ 
T^\sM_{~\sA\sB} \, \nt_\sC \, \xi_\sM \,\,.  
\ee
This equation is not entirely similar to Eq. (\ref{R-intcond}) above, 
because of the second term on the right hand side. However, one may still 
use this as the integrability criterion for the Killing vectors in a 
space-time with torsion. The reason is that all the arguments that follow 
in GR, after getting Eq. (\ref{R-intcond}), would be the same here as well, 
once the Eq. (\ref{U-intcond}) is set. Given the values of $\xi_\sN$ and 
$\nt_\sM \xi_\sN$ at some point $X$, Eq. (\ref{U-intcond}) gives the second 
derivative, and successive differentiations of Eq. (\ref{U-intcond}) yield 
the corresponding higher derivatives of $\xi_\sN$ at $X$. Consequently, a 
particular Killing vector $\xi_\sN^{(q)} (x)$, is only linearly dependent 
on the initial values $\xi_\sN^{(q)} (X)$ and $\nt_\sM \xi_\sN^{(q)} 
\vert_{x=X}$:    
\be \la{U-Killexpand}
\xi_\sN^{(q)} (x) ~=~ \cA_\sN^{~\sM} (x, X) \, \xi_\sM^{(q)} (X) ~+~ 
\cB_\sN^{~\sL\sM} (x, X) \, \nt_\sL \, \xi_\sM^{(q)} \vert_{x=X} \,\,,
\ee
where the coefficients $\cA_\sN^{~\sM}$ and $\cB_\sN^{~\sL\sM}$ depend on 
the metric and the torsion, and are the same for all Killing vectors. Now, 
in an $n$-dimensional space, for every $q$, there can be at most $n$ 
independent quantities $\xi_\sN^{(q)} (X)$ and $n (n-1)/2$ independent 
quantities $\nt_\sM \xi_\sN^{(q)} \vert_{x=X}$ (by virtue of the Killing 
equation (\ref{U-Killeq})). So, any linearly independent set of Killing 
vectors, in $n$ dimensions, can consist of a maximum number of ~$n \,+\, 
n (n-1)/2 \,=\, n (n+1)/2$~ of such vectors. Accordingly, one may say that 
an $n$-dimensional space which admits all of the $n (n+1)/2$ independent 
Killing vectors is maximally symmetric in presence of torsion, as long as 
the torsion tensor satisfies the above two conditions (\ref{tor-Killcond}) 
and (\ref{tor-intcond}).     

As to the maximal symmety of subspaces of a bulk space-time involving 
torsion, the arguments are similar to the above. However, one requires 
the prior assumption that the bulk metric is decomposed exactly in the 
same way as in GR, viz. the Eq. (\ref{bulkmet}) holds. We of course 
make this assumption here, without attempting the rigorous proof of 
Eq. (\ref{bulkmet}) in presence of torsion\symbolfootnote[5]{For the 
proof of Eq. (\ref{bulkmet}) in a torsionless scenario, see \cite{wein}.}.  

Let us now turn our attention to the conditions (\ref{tor-Killcond}) and 
(\ref{tor-intcond}), and see to what extent they can constrain the torsion 
tensor components. In the following two subsections, we shall treat 
separately the cases of (i) the entire (bulk) space-time being maximally 
symmetric, and of (ii) the maximally symmetric subspaces of the bulk.

\subsection{Constraints on torsion due to the maximal symmetry of the bulk 
\la{sec:ms-bulk}}

\no
When the $d$-dimensional bulk space-time admits the maximum number of 
$d(d+1)/2$ independent Killing vectors $\xi^\sA$, which are of course 
arbitrary, the condition (\ref{tor-Killcond}) implies that the torsion 
tensor should be antisymmetric in the first two indices, i.e. 
$T_{\sA\sB\sC} = - T_{\sB\sA\sC}$. But the torsion tensor is 
antisymmetric in its last two indices as well. So one infers that 
it should be completely antisymmetric: $T_{\sA\sB\sC} = T_{[\sA\sB\sC]}$. 
Moreover, recalling that maximal symmetry implies the Killing vectors 
$\xi_\sA$ be chosen such that at a given point $X$, they vanish and 
their covariant derivatives $\nt_\sB \xi_\sA$ are arbitrary (and of 
course antisymmetric because of the Killing equation (\ref{U-Killeq})). 
So, at $X$, the left hand side of the condition (\ref{tor-intcond})
could be made to vanish, which means that on the right hand side the 
coefficient of the antisymmetric tensor $\nt_\sN \xi_\sM$ is symmetric 
in $N$ and $M$:
\be \la{tor-intcond1a}
\d^\sN_\sC \, T^\sM_{~ \sA\sB} ~+~ \d^\sN_\sA \, T^\sM_{~ \sB\sC} ~+~ 
\d^\sN_\sB \, T^\sM_{~ \sC\sA} ~=~  \d^\sM_\sC \, T^\sN_{~ \sA\sB} ~+~ 
\d^\sM_\sA \, T^\sN_{~ \sB\sC} ~+~ \d^\sM_\sB \, T^\sN_{~ \sC\sA} \,\,.
\ee
This holds everywhere as the point $X$ is also arbitrary. 
Contraction of $M$ with $C$ yields
\be \la{tor-intcond1b}
\le(d - 3\ri) T^\sN_{~ \sA\sB} ~=~  \d^\sN_\sA \, T_\sB ~-~ 
\d^\sN_\sB \, T_\sA \,\,.
\ee
As $T_{\sA\sB\sC}$ is completely antisymmetric, its trace $T_\sA = 
T^\sB_{~\sA\sB}$ is zero, i.e. the right hand side of Eq. 
(\ref{tor-intcond1b}) vanishes. Therefore, $T_{\sA\sB\sC}$ could be 
non-vanishing only when the bulk has dimensionality $d = 3$. Moreover,
$T_{\sA\sB\sC}$ is a constant since it cannot depend on any of the 
maximally symmetric bulk coordinates.

This is of course a known result, which has been demonstrated in 
different contexts previously \cite{bloom,dgssg,mult}. We, in this 
section, have taken the route of \cite{dgssg} in which the authors 
have made a comprehensive study of maximal symmetry in presence of 
a completely antisymmetric torsion, i.e. the one for which Eq. 
(\ref{tor-Killcond}) is satisfied automatically. However, the 
condition (\ref{tor-intcond}) which we find here is not the same as 
the conditions imposed in \cite{dgssg} on the completely antisymmetric 
torsion due to the maximal symmetry of the entire space-time. 
In fact, the authors in \cite{dgssg} have demanded that the second 
term on the right hand side of Eq. (\ref{U-commvec}) should 
vanish altogether, and also the part $\bR^\sM_{~\sC\sA\sB}$ of 
$\Rt^\sM_{~\sC\sA\sB}$ should have the same cyclicity property as 
exhibited by the Riemann curvature tensor $R^\sM_{~\sC\sA\sB}$ (viz. 
Eq. (\ref{R-cycl})). But these restrictions on torsion are not essential 
for obtaining Eq. (\ref{U-intcond1}) and carrying out the analysis 
thereafter in a similar manner as in GR. What is sufficient is the 
condition (\ref{tor-intcond}) that we have here.

\subsection{Constraints on torsion in maximally symmetric subspaces of 
the bulk \la{sec:ms-sub}}

\no
When the $d$-dimensional bulk space-time is not maximally symmetric on 
the whole, but can be decomposed into subspaces of dimensionality say 
$n \,(< d)$ which are maximally symmetric, the torsion tensor can be 
constrained in the following way. 

We adopt the same notations and conventions as in section \ref{sec:App1}, and
have the only surviving Killing vectors to be $\xi_{\bq}$, with $\bq$ taking 
$n$ values corresponding to the coordinate labels of an $n$-dimensional
maximally symmetric  sub-manifold $\bcM$. All the Killing vectors of the 
quotient space $\cM/\bcM$ are identically zero. Also, as mentioned above, we 
assume that the metric $g_{\sA\sB}$ of the $d$-dimensional bulk manifold $\cM$ 
is decomposed as in Eq. (\ref{bulkmet}), so that the elements $g_{\ba a}$ do 
not exist, and the Killing vectors $\xi_{\bq}$ are functions only of the 
coordinates $\bu^{\ba}$ of $\bcM$, i.e. $\xi_{\bq} = \xi_{\bq} (\bu)$. The 
Killing equation is now required to be given by
\be \la{U-Killeq2}
\nt_{\ba} \, \xi_{\bb} ~+~ \nt_{\bb} \, \xi_{\ba} ~=~ 0 \,\,,
\ee
as a generalization of $\nab_{\ba} \xi_{\bb} + \nab_{\bb} \xi_{\ba} = 0$ when
torsion had not been there. Thus, instead of Eq. (\ref{tor-Killcond}) we have 
the condition
\be \la{tor-Killcond2}
\le(T_{\ba\bb\bc} ~+~ T_{\bb\ba\bc}\ri) \xi^{\bc} ~=~ 0 \,\,.
\ee
Moreover, among the quantities $\nt_\sA \, \xi_{\bb}$, the existent ones are 
$\nt_{\ba} \, \xi_{\bb}$. So Eq. (\ref{tor-intcond}) becomes
\be \la{tor-intcond2}
\le(\bR^{\bq}_{~\sA\sB\sC} + \bR^{\bq}_{~\sB\sC\sA} + 
\bR^{\bq}_{~\sC\sA\sB}\ri) \xi_{\bq} = 
- \le(\d^{\bp}_{\sC} \, T^{\bq}_{~\sA\sB} + 
\d^{\bp}_{\sA} \, T^{\bq}_{~\sB\sC} + 
\d^{\bp}_{\sB} \, T^{\bq}_{~\sC\sA}\ri) \nt_{\bp} \, \xi_{\bq} \,. 
\ee
Once again, we can make the choice that at a given point $\bu = \bU$, 
$\xi_{\bq}$ vanishes and $\nt_{\bp} \xi_{\bq}$ is an arbitrary antisymmetric 
tensor. Therefore, the coefficient of $\nt_{\bp} \xi_{\bq}$ is symmetric at 
$\bU$ (and of course, everywhere, since the point $\bU$ is also arbitrary):
\be \la{tor-intcond2a}
\d^{\bp}_{\sC} \, T^{\bq}_{~\sA\sB} ~+~ 
\d^{\bp}_{\sA} \, T^{\bq}_{~\sB\sC} ~+~ 
\d^{\bp}_{\sB} \, T^{\bq}_{~\sC\sA} ~=~ 
\d^{\bq}_{\sC} \, T^{\bp}_{~\sA\sB} ~+~ 
\d^{\bq}_{\sA} \, T^{\bp}_{~\sB\sC} ~+~ 
\d^{\bq}_{\sB} \, T^{\bp}_{~\sC\sA} \,\,.
\ee
This is analogous, but not identical, to Eq. (\ref{torinv2}) for the 
form-invariance of torsion under maximal symmetry defined conventionally 
in scheme I (see section \ref{sec:App1}). Accordingly, the constraints 
on torsion here may differ in general from those obtained in section 
\ref{sec:App1}. Let us work out these constraints by applying the Eqs. 
(\ref{tor-Killcond2}) and (\ref{tor-intcond2a}) on each of the six 
independent torsion tensor components $\,T_{\ba \bb \bc} \,, \, T_{\ba \bb a} 
\,, \, T_{\ba a b} \,, \, T_{a b \ba} \,, \, T_{a \ba \bb} \,$ and $\, 
T_{a b c}$.
\bi 
\item For $T_{\ba \bb \bc}$:  The condition (\ref{tor-Killcond2}) implies 
antisymmetry in the first two indices (as $\xi^{\bc}$ is arbitrary). But 
$T_{\ba \bb \bc}$ is antisymmetric in the last two indices as well. So, we 
infer that it should be completely antisymmetric, i.e.  $T_{\ba \bb \bc} = 
T_{[\ba \bb \bc]}$. Moreover, the condition (\ref{tor-intcond2a}) suggests 
the dimensionality of the maximally symmetric submanifold to be $n=3$, in 
the same way as in the previous subsection. Hence, one can express
\be \la{T-cond1}
T_{\ba \bb \bc} ~=~ \e_{\ba \bb \bc} \, \b (v) \,\,, 
\qquad (n = 3 ~ \mbox{only}) \,\,,
\ee
where $\b (v)$ is a pseudo-scalar.
 
\item For $T_{\ba \bb a}$: The condition (\ref{tor-Killcond2}) is not 
applicable, whereas the condition (\ref{tor-intcond2a}) gives
\be \la{T-cond2a}
\d^{\bp}_{\ba} \, T^{\bq}_{~\,\bb a} ~+~ 
\d^{\bp}_{\bb} \, T^{\bq}_{~a \ba} ~=~ 
\d^{\bq}_{\ba} \, T^{\bp}_{~\,\bb a} ~+~ 
\d^{\bq}_{\bb} \, T^{\bp}_{~a \ba}  \,\,.
\ee
Contracting $\bp$ with $\ba$, and using the fact that $T_{\ba \bb a} = 
- T_{\ba a \bb}$, we get
\be \la{T-cond2}
T_{\ba\bb a} ~=~ - \, T_{\ba a \bb} ~=~ 
\fr 1 {n-2} \, \d_{\ba\bb} \, \a_a (v) \,\,; 
\qquad (\forall \, n \neq 2) \,\,,
\ee
where $\a_a = T^{\ba}_{\,\,\, a \ba}$. For $n=2$ not much can be said about 
$T_{\ba\bb a}$ except that it is trace-free ($\a_a = 0$), i.e. at best
we can express
\be \la{T-cond2b}
T_{\ba\bb a} ~=~ T_{[\ba\bb a]} (v) ~+~ Q_{\ba\bb a} (v) \,\,;  
\qquad (n = 2) \,\,,
\ee
where $Q_{\ba\bb a}$ is the (pseudo-)tracefree irreducible mode of 
torsion (see the Appendix for general definition) which satisfies the
condition $Q_{\ba\bb a} + Q_{\bb a \ba} + Q_{a \ba\bb} = 0$.

\item For $T_{\ba a b}$: The condition (\ref{tor-Killcond2}) is again 
not applicable, whereas (\ref{tor-intcond2a}) gives
\be \la{T-cond3}
T_{\ba a b} ~=~ 0 \,\,, \qquad (\forall \, n \geq 2) \,\,.
\ee

\item For the rest ($\, T_{a b \ba} \,, \, T_{a \ba \bb} \,$ and 
$\, T_{a b c}$): The above conditions (\ref{tor-Killcond2}) and 
(\ref{tor-intcond2a}) yield nothing, however, once again the component
$T_{a b c}$ can be expressed as in Eq. (\ref{fincond6}), because of the 
antisymmetry in its last two indices, and of course due to fact that the 
indices $a, b, c$ are of the $(d-n)$-dimensional quotient space $\cM/\bcM$.
\ei
We thus see that not all types of components of torsion could be restricted
in this scheme\symbolfootnote[6]{It should also be noted that following an
analysis similar to that in sec. \ref{sec:App1}, one can constrain some of
the components of the torsion irreducible modes. However, for brevity, we
are not showing them here.}. In the next section, we shall compare these 
components with those allowed by the scheme I, resorting to some particular 
cases.

\section{A comparison between torsion components allowed in the two schemes
\la{sec:Compare}}

\no
From the analysis for the schemes I and II in the previous two sections,
we observe:
\bi 
\item If the entire bulk manifold $\cM$ is maximally symmetric, then both
the schemes allow for a totally antisymmetric torsion $T_{\sA\sB\sC} = 
T_{[\sA\sB\sC]}$, provided the dimension of the bulk is $d = 3$, so that 
torsion is determined by only one constant parameter.
\item If instead, the maximal symmetry is exhibited only in a submanifold 
$\bcM$, then 
\bi
\item For a totally antisymmetric torsion, the outcome of the two schemes 
are again the same. Torsion is non-vanishing if the dimension of the 
submanifold $\bcM$ is either $n = 2$ or $n = 3$. Whereas for $n=2$ the
only surviving component is $T_{\ba\bb a}$, for $n=3$ the only torsion 
degree of freedom (DoF) is a pseudo-scalar $\b$ which is a function of the 
coordinates of the quotient space $\cM/\bcM$.
\item For a generic torsion (antisymmetric only in a pair of indices) however,
the results of the two schemes differ in general. Whereas scheme I constrains
all types of independent torsion components except one ($T_{abc}$), scheme II 
can at most restrict three types $T_{\ba\bb\bc}, T_{\ba\bb a}$ and 
$T_{\ba a b}$.  
\ei
\ei
Let us now consider, as illustrations, some particular cases of 
physical importance. 

\subsection{Relevant scenarios in four dimensions} 

\no 
$\cM$ is the bulk manifold of dimension $d=4$, with coordinates 
$x^A := x^0, x^1, x^2, x^3$ (i.e. the bulk indices $A, B, \dots$
run over $0, 1, 2, 3$). We have the following scenarios, for which
the allowed torsion tensor components in the schemes I and II are
shown in Table 1:

\bigskip
\no 
(i) $\bcM$: Maximally symmetric submanifold of dimension $n=2$,  
coordinates $\bu^{\ba} := x^2, x^3$ say (i.e. the indices $\ba, 
\bb, \dots = 2, 3$). 

\vskip 0.025in
\no
$\cM/\bcM$: Quotient space of dimension $d-n=2$, coordinates 
$v^a := x^0, x^1$ (i.e. the indices $a, b, \dots = 0, 1$).

\vskip 0.025in
\no
Example: Spherically symmetric space-time ($x^0, x^1 = t, r$ and 
$x^2, x^3 = \vth, \vph$).

\bigskip
\no 
(ii)
$\bcM$: Maximally symmetric submanifold of dimension $n=3$, 
coordinates $\bu^{\ba} := x^1, x^2, x^3$ say (indices $\ba, \bb, \dots 
= 1, 2, 3$). 

\vskip 0.025in
\no
$\cM/\bcM$: Quotient space of dimension $d-n=1$, coordinate 
$v^a := x^0$ (indices $a, b, \dots = 0$).

\vskip 0.025in
\no
Example: Homogeneous and isotropic 
space-time ($x^0 = t$ and $x^1, x^2, x^3 = r, \vth, \vph$). 
%
%%%%%%%%%%%%%%%%%%%%%%%%%%%%%%%%%%% TABLE STARTS %%%%%%%%%%%%%
\begin{table*}[!htb]
\la{table1}
\begin{tabular}{|c||c|c||c|c|}
\hline
Submanifold&\multicolumn{2}{c|}{Scheme I}&\multicolumn{2}{c|}{Scheme II} \\
\cline{2-5}
dimensionality&Allowed components&DoF&Allowed components&DoF \\
\hline\hline 
&\ul{{\small Type} $T_{\ba\bb a}$~:}& 
&\ul{{\small Type} $T_{\ba\bb a}$~:}&  \\
&$T_{220} = T_{330} = - \fr 1 2 \a_{_0} \,,$& & & \\
{\Large $n = 2$}&$T_{221} = T_{331} 
= - \fr 1 2 \a_{_1} \,,$
&$4$&$T_{230} \,,~ T_{320} \,,$&$4$ \\
&$T_{230} \!=\! T_{[230]} ,\, T_{231} \!=\! T_{[231]} .$ & &
$T_{231} \,,~ T_{321} \,.$& \\
%
% & & & & \\
%
\cline{2-5}
&\ul{{\small Type} $T_{a b c}$~:}& 
&\ul{{\small Type} $T_{a b c}$~:}& \\
{\small $\le[a, b, \dots = 0,1 \ri]$}
&$T_{001} \,,~ T_{110} \,.$&$2$ 
&$T_{001} \,,~ T_{110} \,.$&$2$ \\
%
% & & & & \\
%
\cline{2-5}
& & &\ul{{\small Type} $T_{a \ba\bb}$~:}& \\
{\small $\le[\ba, \bb, \dots = 2,3 \ri]$}
& & &$T_{023} \,,~ T_{123} \,.$&$2$ \\
%
% & & & & \\
%
\cline{2-5}
& & &\ul{{\small Type} $T_{a b \ba}$~:}& \\
& & &$T_{002} ,\, T_{003} ,\, T_{112} ,\, T_{113} ,$&$8$ \\
& & &$T_{012} ,\, T_{102} ,\, T_{013} ,\, T_{103} .$ & \\
 & & & & \\
\hline\hline
&\ul{{\small Type} $T_{\ba\bb\bc}$~:}& 
&\ul{{\small Type} $T_{\ba\bb\bc}$~:}& \\
{\Large $n = 3$}&$T_{123} = T_{[123]} = \b $&$1$&
$T_{123} = T_{[123]} = \b $&$1$ \\
%  
% & & & & \\
%
\cline{2-5}
&\ul{{\small Type} $T_{\ba\bb a}$~:}& 
&\ul{{\small Type} $T_{\ba\bb a}$~:}& \\
{\small $\le[a, b, \dots = 0 \ri]$}&$T_{110} = T_{220} = T_{330} = - \fr 1 3 \a_{_0} \,.$
&$1$&$T_{110} = T_{220} = T_{330} = \a_{_0} \,.$&$1$ \\
%
% &$T_{330} = - \fr 1 3 \a_{_0} \,.$& &$T_{330} = \a_{_0} \,.$& \\
%
% & & & & \\
%
\cline{2-5}
& & &\ul{{\small Type} $T_{a b \ba}$~:}& \\
{\small $\le[\ba, \bb, \dots = 1,2,3 \ri]$}
& & &$T_{001}\,,~ T_{002}\,,~ T_{003} \,.$&$3$ \\
 & & & & \\
\hline
\end{tabular}
\caption{\it A comparison of the allowed torsion components in the two 
schemes, for a given manifold of dimension $d=4$ with a maximally
symmetric submanifold of dimension $n = 2, 3$. All the components are in 
general functions of the $(d-n)$-dimensional quotient space coordinates
(i.e. of $x^0,x^1$ for $n=2$, and of $x^0$ for $n=3$).}
\end{table*}
%%%%%%%%%%%%%%%%%%%%%%%%%%%%%%%%%%% TABLE ENDS %%%%%%%%%%%%%
%

\no 
We observe that the torsion DoF are in general different in the two
schemes. In fact, scheme II allows more torsion DoF than scheme I for 
both the cases $n=2$ and $n=3$. Even when the number of DoF for the
components of a particular type are the same in the two schemes, the
components themselves are different. For instance, there are four 
allowed components of the type $T_{\ba\bb a}$ for $n=2$ in either
scheme, but these components are not the same. For $n=3$ also, the
allowed components of the type $T_{\ba\bb a}$ are $T_{110} = T_{220}
= T_{330}$ in both the schemes, but the values of these components
are different. So the schemes I and II are not in general equivalent.
However, there is an interesting point to note. If the torsion tensor 
is completely antisymmetric in its indices, then for $n=2$ both the
schemes allow two DoF and the components are also the same, viz. 
$T_{230}$ and $T_{231}$. Similarly, for $n=3$ we have only one torsion
DoF, viz. the pseudo-scalar $\b$ (which is a function of $x^0$), in 
both the schemes. Thus, with the additional property of complete
antisymmetry in the indices, the torsion tensor apparently does not
distinguish between the schemes I and II. The reason for this could be
traced to the fact that a completely antisymmetric torsion covariantly
preserves the Killing equation (see eq. (\ref{tor-Killcond})), although
it alters the equations relevant for the integrability of the latter.

\subsection{A higher dimensional example} 

\no
For simplicity, let us consider the following:
\bi 
\item 
$\cM$: Bulk manifold of dimension $d=5$, coordinates $x^A := x^0, x^1, x^2, 
x^3, y$ ~(indices $A, B, \dots = 0, 1, 2, 3, y$).

\item
$\bcM$: Maximally symmetric submanifold of dimension $n=4$, coordinates 
$\bu^{\ba} := x^0, x^1, x^2, x^3$~ (indices $\ba, \bb, \dots = 0, 1, 2, 3$). 

\item 
$\cM/\bcM$: Quotient space of dimension $d-n=1$, coordinate 
$v^a := y$~ (indices $a, b, \dots = y$).
\ei
The `extra' (fifth) coordinate $y$ is presumably compact, and the chosen
scheme of compactification may be, for example, the Randall-Sundrum (RS) 
$S^1/Z_2$ orbifolding \cite{rs}. The minimal version of the two-brane RS 
model assumes the bulk geometry to be anti-de Sitter, with the hidden and 
the visible branes located at two orbifold fixed points $y = 0$ and $y = 
r_c \pi$ respectively, $r_c$ being the brane separation. We can consider 
this to be true here as well, alongwith the supposition that torsion 
co-exists with gravity in the bulk. The RS five dimensional line element  
\be \la{RS-met}
ds^2 ~=~ e^{- 2 \s(y)} \, \eta_{\ba\bb} (\bu) \, d\bu^{\ba} \, d\bu^{\bb} 
~+~ dy^2 \,\,,
\ee
describes a non-factorizable geometry with an exponential warping, given by
the {\em warp factor} $\s (y)$, over a four dimensional flat (Minkowski) 
metric $\eta_{\ba\bb} (\bu)$. One can see that Eq. (\ref{RS-met}) shows
a structural breakup similar to that in Eq. (\ref{bulkmet}) for the metric 
of any given manifold (say of dimension $d=5$) with a maximally symmetric 
submanifold (say of dimension $n=4$). The objective of the RS model is to 
provide a resolution to the well known fine-tuning problem of the Higgs 
mass against radiative corrections due to the gauge hierarchy. In the usual
(torsionless) picture, the solution for the warp factor $\s (y)$ turns out
to be linear in $|y|$ \cite{rs}. Therefore, applying the boundary conditions 
one finds that the four dimensional Planck mass $M_p$ is related to the five 
dimensional Planck mass $M$ as
\be \la{eff-Pl}
M_p^2 ~=~ \fr{M^3}{k} \le(1 - e^{- 2 k r_c \pi}\ri) \,, \qquad \le[k 
\sim M\ri] \,.
\ee
That is, the Planck-electroweak hierarchy could effectively be made to 
subside on the visible brane (our observable four dimensional world) 
by appropriate adjustment of the parameters in the exponential factor 
$e^{- 2 k r_c \pi}$. In fact, setting $k r_c \simeq 12$ achieves the 
desired stabilization of the Higgs mass. In presence of the bulk torsion 
however, the solution for the warp factor would be altered. That is 
torsion would backreact on the RS warping. Such a backreaction would
have its immediate effect on the stability of the RS model, in the sense
that the existent torsion DoF would describe the dynamics of the {\em
radion}, i.e. the field which governs the fluctuations in the brane
separation $r_c$ \cite{rad}. Now, the torsion tensor components that can 
take part in the backreaction, and as such in the radion stabilization, 
are shown in Table 2 for the schemes I and II. 
%
%%%%%%%%%%%%%%%%%%%%%%%%%%%%%%%%%%% TABLE STARTS %%%%%%%%%%%%%
\begin{table*}[!htb]
\la{table2}
\begin{tabular}{|c||c|c||c|c|}
\hline
Submanifold&\multicolumn{2}{c|}{Scheme I}&\multicolumn{2}{c|}{Scheme II} \\
\cline{2-5}
dimensionality&Allowed components&DoF&Allowed components&DoF \\
\hline\hline 
&\ul{{\small Type} $T_{\ba\bb a}$~:}& 
&\ul{{\small Type} $T_{\ba\bb a}$~:}& \\
{\Large $n = 4$}&$T_{00y} = T_{11y}= T_{22y}$&$1$&
$T_{00y} = T_{11y} = T_{22y}$ &$1$ \\  
&$= T_{33y} = - \fr 1 4 \a_{_y} (y) \,.$& & 
$= T_{33y} = \fr 1 2 \a_{_y} (y) \,.$ & \\
& & & & \\
\cline{2-5}
& & &\ul{{\small Type} $T_{a b \ba}$~:}& \\
{\small $\le[a, b, \dots = y \ri]$}
& & &$T_{yy0}\,,~ T_{yy1}\,,~ T_{yy2}\,,~ T_{yy3}\,.$&$4$ \\
& & & & \\
\cline{2-5}
& & &\ul{{\small Type} $T_{a \ba\bb}$~:}& \\
{\small $\le[\ba, \bb, \dots = 0,1,2,3 \ri]$}
& & &$T_{y01}\,,~ T_{y02}\,,~ T_{y03}\,,~$ &$6$ \\
& & &$T_{y12}\,,~ T_{y23}\,,~ T_{y31}\,.$ & \\
\hline
\end{tabular}
\caption{\it A comparison of the allowed torsion components in the two 
schemes, for a given manifold of dimension $d = 5$ with a maximally
symmetric submanifold of dimension $n = 4$. All the components are 
in general functions of the  coordinate $y$.}
\end{table*}
%%%%%%%%%%%%%%%%%%%%%%%%%%%%%%%%%%% TABLE ENDS %%%%%%%%%%%%%
%

\no
We clearly see that neither of the schemes allow for a completely
antisymmetric torsion. However, for a generic torsion (antisymmetric 
in the last two indices), one DoF is allowed in scheme I allows whereas 
scheme II allows as many as $1+4+6 = 11$ DoF. So, the warping (and hence 
the overall aspect of, for e.g., the radion stabilization) is expected 
to be affected in different ways for the two 
schemes\symbolfootnote[3]{It is worth mentioning here that one should 
not get confused with the result in Ref. \cite{bmssgsen} that the 
massless mode arising from a bulk torsion field is heavily suppressed 
by the exponential RS warping in the visible brane. The authors in 
\cite{bmssgsen} assumed torsion to be completely antisymmetric (being 
induced by the Kalb-Ramond field in a string-inspired picture) and did 
not consider restricting it on account of maximal symmetry exhibited by 
the four dimensional flat (Minkowski) submanifold.}.

\section{Conclusions  \la{sec:Concl}}

\no 
We have thus addressed certain conceptual issues related to the basic
understanding of symmetries of space-times admitting torsion. In 
particular, we have concentrated on determining the independent torsion
degrees of freedom that are allowed for the preservation of maximal
symmetry of either the entire bulk manifold or of its subspaces. This
is of importance in implicating torsion's role in a variety of physical
scenarios and in a number of observable phenomena. In fact, one may
realize that the whole concept of maximal symmetry deserves a proper 
clarification in presence of torsion. This has been addressed in a few
earlier works \cite{tsim,zec,boem,bloom,dgssg,mult} either in a direct
way or in some specific contexts. The ideas put forward may be summed 
up into two different schemes of implementing the symmetry concepts in
space-times with torsion. In the first of these (scheme I), the maximal
symmetry of (sub)spaces is supposedly being sensed in the usual way (as
in GR), the only demand is that torsion should be form-invariant under
the infinitesimal isometries of the metric of such (sub)spaces. We have
made a careful examination of the torsion components, thus constrained,
in all possible scenarios. Scheme II is more robust as this requires a
complete covariant generalization of the GR conception of maximal symmetry
in presence of torsion. Under strict enforcement of the minimal coupling 
prescription ($\nab \lra \nt$), such a generalization amounts to the 
preservation of the Killing equation (which is in general modified by 
torsion), and the essential outcome of the integrability of the same.
Unlike in ref. \cite{dgssg}, we have looked for the conditions of absolute
necessity to be thus imposed on torsion. Such conditions enabled us to
identify the allowed independent torsion components (for the scheme II), 
which we have compared with those that are allowed in scheme I. Although
these components are in general different, we find that in the special 
cases of a maximally symmetric bulk manifold or(and) a completely 
antisymmetric torsion, they are identical in the two schemes. In fact, 
if the entire bulk is maximally symmetric only a completely antisymmetric 
torsion is allowed, and that also when the bulk dimensionality is only 
$d=3$. Thus, at least for the completely antisymmetric torsion, we can 
uniquely identify its components that can preserve the maximal symmetry
of given (sub)spaces.
      
We have made illustrations of particular cases of physical interest in 
the context of both four and higher dimensional theories. For the four 
dimensional bulk space-time, we have explored the relevant cases, viz. 
maximally symmetric submanifold of dimensionality $n=2$ and $3$. These 
cases correspond respectively to, for e.g., a spherically symmetric 
space-time and a homogeneous and isotropic (cosmological) space-time. 
So our analysis may be useful in examining (say) the viability of the 
spherically symmetric black-hole solutions in presence of torsion, or the 
role of torsion in the context of cosmological inflation or the problem 
of dark energy. As to the higher dimensional example, we have considered 
a five dimensional bulk manifold (which admits torsion) with a maximally 
symmetric four dimensional submanifold. The general structure of the bulk 
metric has resemblance with that of the Randall-Sundrum (RS) two-brane 
model \cite{rs}, which aims to resolve the fine-tuning of the Higgs mass 
due to the Planck-electroweak hierarchy. However, the bulk torsion is 
expected to backreact on the RS warp factor, and we have actually been 
able to figure out which of the torsion components would be responsible 
for that. Such a backreaction can have its significance in, for e.g., the
the {\em radion} stabilization \cite{rad}. So our analysis of determining 
the allowed torsion components would enable one to examine the role of 
torsion (if at all conceivable) in the stability of the RS brane-world. 

Some issues remain open in the context of this paper. 
Firstly, one has to prove rigorously whether the general
structural breakup of the metric, viz. Eq. (\ref{bulkmet}), is indeed
valid if one adopts the line of approach of scheme II. That should be
a consistency check for the scheme II. Secondly, it has to be verified
whether the scheme II at all provides a unique way of integrating the
Killing equation in space-times with torsion. That is to say, whether it 
is absolutely necessary in a covariant generalization in presence of 
torsion that one should maintain the exact GR analogy at every crucial 
step. Thirdly, what would happen with relevance to say the non-minimal 
coupling of the torsion modes to scalar or tensor fields? What would be
the consequential effects in cosmology, astrophysics, brane-world 
scenarios, string-motivated phenomenology? Works are under way to 
explore some of these issues \cite{ss1,ssasb} which we hope to report 
soon.

\section*{Acknowledgement}

\no 
SS acknowledges illuminating discussions with Soumitra SenGupta. The 
work of ASB is supported by the Council of Scientific and Industrial 
Research (CSIR), Government of India. The authors are also grateful
to the anonymous referee for useful comments and suggestions.

%\newpage

%%%%%%%%%%%%%%%%%%%%%%%%%%%%%%%%%%%%%%%%%%%%%%%%%
%%%%%%%%%%%%%%%%%%%%%%% APPENDIX
%%%%%%%%%%%%%%%%%%%%%%%%%%%%%%%%%%%%%%%%%%%%%%%%%

%\appendix
%\noappendicestocpagenum
%\appendixpage
%\addappheadtotoc

%\begin{appendices}
%\setcounter{secnumdepth}{0}
  
  \renewcommand{\theequation}{A-\arabic{equation}}
  % redefine the command that creates the equation no.
  \setcounter{equation}{0}  % reset counter 

\section*{Appendix: General Characteristics of Riemann-Cartan space-time 
\la{sec:RC}}

\no 
In a $d$-dimensional Riemannian space-time (R$_d$), the formulation of General 
Relativity (GR) is based on two essential requirements: 
\ben[(i)]
\item symmetry of the affine connection $\C^\sA_{~\sB\sC} = \C^\sA_{~\sC\sB}$, 
and 
\item metricity of the covariant derivative $\nab_\sM \, g_{\sA\sB} = 0$, where 
$g_{\sA\sB}$ is the metric tensor.
\een
By virtue of these, the standard expression of the covariant derivative of any 
arbitrary tensor $V^{\sA \dots}_{~~~~\sB \dots}$, viz.
\be \la{covdiff}
\nab_\sM \,V^{\sA \dots}_{~~~~\sB \dots} :=~ 
\pa_\sM \, V^{\sA \dots}_{~~~~\sB \dots} + 
\C^\sA_{~\sN\sM} V^{\sN \dots}_{~~~~\sB \dots} + \cdots 
- \C^\sL_{~\sB\sM} V^{\sA \dots}_{~~~~\sL \dots} - \cdots 
\ee
leads to a unique solution for $\C^\sA_{~\sB\sC}$ in the form of the Christoffel 
symbol
\be \la{Chr}
\C^\sA_{~\sB\sC} ~:=~ \fr 1 2 \, g^{\sA\sM} \le(\pa_\sC \, g_{\sM\sB} ~+~ 
\pa_\sB \, g_{\sC\sM} ~-~ \pa_\sM \, g_{\sB\sC}\ri) \,.
\ee
However, referring back to Eq. (\ref{covdiff}) we see that its left hand side 
transforms as a tensor even when one adds an arbitrary tensor $K^\sA_{~\sB\sC}$ 
to any given connection $\C^\sA_{~\sB\sC}$ (which itself is of course not a 
tensor). That is, there is an ambiguity in the definition of the affine 
connection right from the beginning, and only the above requirements make the 
connection uniquely determined in GR. Relaxation of even the first requirement 
(i.e. symmetry property of the connection) leads to the formulation of one of
the most simplest and natural modifications of GR, in the $d$-dimensional 
Riemann-Cartan (U$_d$) space-time \cite{hehl,akr,sab,shap}. Such a space-time 
is characterized by an asymmetric affine connection $\Ct^\sA_{~\sB\sC}$ which 
is related to the Christoffel symbol as
\be \la{RC-conn}
\Ct^\sA_{~\sB\sC} ~=~ \C^\sA_{~\sB\sC} ~+~ K^\sA_{~\sB\sC} \,\,.
\ee
Although this new connection is still non-tensorial, the antisymmetrization of 
its last two indices gives rise to a tensor, referred to as {\em torsion}:
\be \la{tor-def}
T^\sA_{~\sB\sC} ~:=~ 2 \, \Ct^\sA_{~[\sB\sC]} ~\equiv~ \Ct^\sA_{~\sB\sC} ~-~ 
\Ct^\sA_{~\sC\sB} \,.
\ee
Moreover, assuming that the metricity condition would still hold for the new 
covariant derivatives $\nt_\sM$ (in terms of $\Ct^\sA_{~\sB\sC}$), i.e. 
$\nt_\sM \, g_{\sA\sB} = 0$, one can express the tensor $K^\sA_{~\sB\sC}$ 
(known as {\em contorsion}) in the form:
\be \la{contor}
K^\sA_{~\sB\sC} ~=~ \fr 1 2 \le(T^\sA_{~\sB\sC} ~-~ T^{~\sA}_{\sB~\sC} ~-~ 
T^{~\sA}_{\sC~\sB}\ri) \,.
\ee
Torsion being antisymmetric in the last two indices, one may verify that the 
contorsion is antisymmetric in the first two indices
\be
T_{\sA\sB\sC} ~=~ T_{\sA[\sB\sC]} \quad, \qquad K_{\sA\sB\sC} ~=~ 
K_{[\sA\sB]\sC} \,\,.
\ee
The torsion tensor can further be decomposed into three irreducible components 
as \cite{shap,neto,capo}:
\be \la{torirrd}
T_{\sA\sB\sC} ~=~ \fr 1 {d-1} \le(g_{\sA\sC} \, T_\sB ~-~ g_{\sA\sB} \, 
T_\sC\ri) \,+~ A_{\sA\sB\sC} ~+~ Q_{\sA\sB\sC} \,\,,
\ee
where
\bi
\item $T_\sC$ is the torsion {\em trace} vector, given by
\be \la{T-tor} 
T_\sC ~:=~ T^{\sA}_{~\,\sC\sA} ~=~ - \, T^{\sA}_{~\,\sA\sC} \,\,.
\ee
\item $A_{\sA\sB\sC}$ is the {\em completely antisymmetric} part of torsion, 
given by
\be \la{A-tor}
A_{\sA\sB\sC} ~:=~ T_{[\sA\sB\sC]} ~\equiv~ \fr 1 3 \le(T_{\sA\sB\sC} ~+~ 
T_{\sB\sC\sA} ~+~ T_{\sC\sA\sB}\ri) \,,
\ee
which in four dimensions ($d=4$) is expressed as $A_{\a\b\c} = \fr 1 6 \, 
\e_{\a\b\c\d} \cA^\d$, ~($\cA^\d$ being the torsion {\em pseudo-trace}). 
\item $Q_{\sA\sB\sC}$ is the {\em (pseudo-)traceless} part of torsion, which 
satisfies the conditions
\be \la{Q-tor}
Q_{\sA\sB\sC} = Q_{\sA[\sB\sC]} \,\,, \quad Q^{\sA}_{~\sC\sA} = 0 \,\,, 
\quad Q_{\sA\sB\sC} + Q_{\sB\sC\sA} + Q_{\sC\sA\sB} = 0 \,.
\ee
\ei
Covariant derivatives in the Riemann-Cartan (U$_d$) space-time are analogous 
to those in the Riemannian (R$_d$) space-time. For any tensor field 
$V^{\sA \dots}_{~~~\sB \dots}$ the U$_d$ covariant derivative is given by
\bea \la{torcovdiff}
\nt_\sM \,V^{\sA \dots}_{~~~~\sB \dots} &:=& \pa_\sM \, 
V^{\sA \dots}_{~~~~\sB \dots} + 
\Ct^\sA_{~\sN\sM} V^{\sN \dots}_{~~~~\sB \dots} + \cdots -
\Ct^\sL_{~\sB\sM} V^{\sA \dots}_{~~~~\sL \dots} - \cdots \nn\\
&=& \nab_\sM \, V^{\sA \dots}_{~~~~\sB \dots} \!+ K^\sA_{~\sN\sM} 
V^{\sN \dots}_{~~~~\sB \dots} \!+ \cdots - 
K^\sL_{~\sB\sM} V^{\sA \dots}_{~~~~\sL \dots} \!- \cdots \,,
\eea
where $\nab_\sM$ denotes the Riemannian covariant derivative (in terms 
of the Christoffel connection). Such a definition is of course based
on the notion of {\em parallel transport} along the distinguished curves 
whose properties are defined solely by the metric or(and) the connection. 
However, in the Riemann-Cartan space-time one has to make the distinction
between two types of such curves, viz. the {\em geodesics} and the {\em 
auto-parallels}. The geodesics, or more appropriately the {\em metric 
geodesics}, are the curves which extremize the infinitesimal separation 
between events, i.e. $ds^2 = g_{\sA\sB} dx^\sA dx^\sB$, and depend only 
on the metric properties of the space-time. The equation of a geodesic is 
the same as that in the Riemannian geometry:
\be \la{geod}
u^\sB \, \nab_\sB \, u^\sA ~=~ \fr{d u^\sA}{d\l} ~+~ \C^\sA_{~\sB\sC} 
\, d u^\sB \, du^\sC ~=~ 0 \,,
\ee
where $\C^\sA_{~\sB\sC}$ is the usual Christoffel connection, and $\l$ 
parameterizes the geodesic, the tangent vectors to which are denoted
by $u^\sA = dx^\sA/d\l$. The parameter $\l$ could be an {\em affine
parameter} when it is fixed upto an affine transformation $\l \lra \l'
= a \l + b$ where $a$ and $b$ are constants and one may choose $\l$
as $s$ (or $\t$), the proper distance (or time) for space-like (or
time-like) geodesics. However, the geodesics in the Riemann-Cartan 
(U$_d$) space-time are not affine (except in some special cases).
For e.g. the functional
\be \la{proptime}
\D \t ~=~ \int_X^Y \, d\l \sq{\le\vert g_{\sA\sB} \, \fr{dx^\sA}{d\l} \,
\fr{dx^\sB}{d\l}\ri\vert}
\ee 
that is extremized by a time-like geodesic passing through the points 
$x^\sA (X)$ and $x^\sA (Y)$, cannot be interpreted as the proper time 
interval between these two points as the functional $\D \t$ is not
invariant under the Cartan transformations, i.e. the transformations 
of the connection involving torsion (see ref. \cite{fneto} for further
clarification). The autoparallels, on the other hand, are the curves
which transport their tangent vectors parallely along themselves. In 
the U$_d$ space-time, the equation of an auto-parallel curve,  
parameterized by $\s$, is given by \cite{sab,fneto}   
\be \la{autop}
\fr{D v^\sA}{D\l} ~=~ v^\sB \, \nt_\sB \, v^\sA ~=~ \fr{d v^\sA}{d\s} 
~+~ \Ct^\sA_{~\sB\sC} \, d v^\sB \, dv^\sC ~=~ 0 \,,
\ee
where $v^\sA = dx^\sA/d\s$ denotes the tangent vector to the 
auto-parallel. The parameter $\s$ is truely affine, since upto an
affine transformation it can be identified with the space-time interval
$s$, given by 
\be \la{propdist}
\le(\fr{ds}{d\s}\ri)^2 ~=~ g_{\sA\sB} \, \fr{dx^\sA}{d\s} \,
\fr{dx^\sB}{d\s} 
\ee
which remains constant under parallel transport along the auto-parallel 
curve in the Riemann-Cartan space-time \cite{fneto}. Hence the auto-parallels
are often referred to as the {\em affine geodesics} and these are the curves
which are of importance in the context of defining the Riemann-Cartan covariant 
derivatives. It is to be noted that the (metric) geodesics coincide with the
auto-parallels, i.e. Eqs. (\ref{geod}) and (\ref{autop}) are identical, in
the case where the torsion tensor is completely antisymmetric, i.e. $T_{\sA\sB\sC}
= A_{\sA\sB\sC} = T_{[\sA\sB\sC]}$.

One may also verify that for any scalar field $\phi$, we have the following
expression
\be \la{commscal}
\le(\nt_\sM \nt_\sN ~-~ \nt_\sN \nt_\sM\ri) \phi ~=~ T^\sL_{~\sM\sN} \, 
\pa_\sL \phi \,\,, 
\ee
which implies that torsion could be sensed even by a scalar field. For a 
vector field $V^\sA$, one has 
\be \la{commvec}
\le(\nt_\sM \nt_\sN ~-~ \nt_\sN \nt_\sM\ri) V^\sA ~=~ T^\sL_{~\sM\sN} \, 
\nt_\sL V^\sA ~+~ \Rt^\sA_{~\sL\sM\sN} V^\sL 
\ee
where $\Rt^\sA_{~\sB\sC\sD}$ is the {\em curvature tensor} defined in the 
Riemann-Cartan (U$_d$) space-time, in analogy with the Riemannian curvature 
tensor $R^\sA_{~\sB\sC\sD}$:
\bea \la{torcurvtens}
\Rt^\sA_{~\sB\sC\sD} &:=& \pa_\sC \, \Ct^\sA_{~\sB\sD} ~-~ 
\pa_\sD \, \Ct^\sA_{~\sB\sC} ~+~ \Ct^\sA_{~\sL\sC} \Ct^\sL_{~\sB\sD} 
~-~ \Ct^\sA_{~\sL\sD} \Ct^\sL_{~\sB\sC} \nn\\
&=& R^\sA_{~\sB\sC\sD} + \nab_\sC \, K^\sA_{~\sB\sD} - 
\nab_\sD \, K^\sA_{~\sB\sC} + K^\sA_{~\sL\sC} K^\sL_{~\sB\sD} - 
K^\sA_{~\sL\sD} K^\sL_{~\sB\sC} \, .
\eea
The U$_d$ analogues of the {\em Ricci tensor} $R_{\sA\sB}$ and the {\em Ricci 
scalar curvature} $R$ of the Riemannian geometry, are given respectively by
\bea 
\la{torRictens}
\Rt_{\sA\sB} &=& \Rt^\sM_{~\sA\sM\sB} ~=~ R_{\sA\sB} + 
\nab_\sM \, K^\sM_{~\sA\sB} - \nab_\sB \, K_\sA + 
K_\sL K^\sL_{~\sA\sB} - K^\sM_{~\sL\sB} K^\sL_{~\sA\sM} \,, \\
\la{torRicscal}
\Rt &=& g^{\sA\sB} \Rt_{\sA\sB} ~=~ R ~-~ 2 \, \nab_\sA \, K^\sA ~-~ 
K_\sA K^\sA ~+~ K_{\sA\sB\sC} K^{\sA\sC\sB} \,,
\eea
where $K_\sA = K^\sB_{~\sA\sB} = T_\sA$ is the trace of the contorsion 
tensor. 

One may note that unlike $R_{\sA\sB}$, the tensor $\Rt_{\sA\sB}$ is not 
symmetric in $A$ and $B$. Moreover, the {\em cyclicity} property of the 
Riemann curvature tensor $R^\sA_{~\sB\sC\sD}$ is not preserved for 
$\Rt^\sA_{~\sB\sC\sD}$ in the U$_d$ space-time
\be \la{Riemcycl}
\Rt^\sA_{~\sB\sC\sD} ~+~ \Rt^\sA_{~\sC\sD\sB} ~+~ \Rt^\sA_{~\sD\sB\sC} 
~\neq~ 0 \,\,.
\ee
In terms of the irreducible torsions components given above, $\Rt$ is 
expressed as \cite{neto}
\be \la{torRicscal1}
\Rt \,=\, R \,-\, 2 \, \nab_\sA \, T^\sA \,- \fr d {2(d-1)} \, T_\sA T^\sA +\, 
\fr 1 4 \, A_{\sA\sB\sC} A^{\sA\sB\sC} \,+\, Q_{\sA\sB\sC} Q^{\sA\sB\sC} \,,
\ee
and this is generally taken as the Lagrangian density for gravity (plus 
torsion) in the Riemann-Cartan space-time.

%\end{appendices}

\end{document}